\documentclass[pra,aps,twocolumn,showpacs,amsmath,amssymb,floatfix,10pt]{revtex4-1}
\input{epsf}

\usepackage{graphicx}% Include figure files
\usepackage{dcolumn}% Align table columns on decimal point
\usepackage{bm}% bold math

\begin{document}
\title{Steady-state entanglement in a double-well Bose-Einstein condensate through coupling 
to a superconducting resonator}
\author{H. T. Ng${}^{1}$ and Shih-I Chu${}^{1,2}$}
\affiliation{${}^{1}$Center for Quantum Science and Engineering, Department of Physics, National Taiwan University, Taipei 10617, Taiwan}
\affiliation{${}^{2}$Department of Chemistry, University of Kansas, Lawrence, Kansas 66045, USA}
\date{\today}

\begin{abstract}
We consider a two-component Bose-Einstein condensate
in a double-well potential, where the atoms are magnetically
coupled to a single-mode of the microwave field inside 
a superconducting resonator. We find that the system has the 
different dark-state subspaces in the strong- and weak-tunneling 
regimes, respectively. In the limit of weak tunnel coupling,   
steady-state entanglement between the two spatially separated 
condensates can be generated by evolving to a mixture of dark states 
via the dissipation of the photon field. We show that 
the entanglement can be faithfully indicated by an entanglement witness.
Long-lived entangled states are useful for quantum information processing
with atom-chip devices.
\end{abstract}


\maketitle

\section{Introduction}
Recently, the realization of a Bose-Einstein condensate (BEC) strongly
coupled to the quantized photon field in an optical cavity has 
been shown \cite{Colombe,Brennecke}.  This paves the way to study 
the interplay of atomic interactions and atom-photon interactions.  
For example, the novel quantum phase transition of a condensate 
coupled to a cavity has been demonstrated \cite{Baumann}.
Strong atom-photon coupling is useful for quantum communications \cite{Duan}
such as the light-matter interface \cite{Colombe,Brennecke}.

Alternatively, strong coupling of ultracold atoms 
to a superconducting resonator has been recently proposed \cite{Henschel}.  
The two long-lived hyperfine states $|e\rangle=|F=2,m_F=1\rangle$ and 
$|g\rangle=|F=1,m_F=-1\rangle$ of ${}^{87}$Rb \cite{Matthews} 
are considered to be magnetically coupled to the microwave field via 
their magnetic dipoles \cite{Henschel,Imamoglu}.  
Since the high-Q superconducting resonator can be fabricated to 
a small mode volume \cite{Wang} and the 
coupling strength can be greatly increased due to the collective 
enhancement \cite{Duan}.  The strong coupling of ultracold atoms 
in the microwave regime can be achieved \cite{Henschel}.  

In this paper, we study a two-component BEC in a double-well 
potential \cite{Ng1}, where all atoms are equally coupled to a single-mode 
of the microwave field inside a superconducting resonator.  
Two weakly linked condensates can be created in a magnetic double-well 
potential on an atom-chip \cite{Schumm,Maussang} or in an optical double-well 
potential \cite{Shin}.  In fact, the tunneling dynamics between the atoms in two wells 
has been recently observed \cite{Albiez,Folling,Trotzky}.  A double-well BEC coupled to an optical
cavity has also been discussed in the literature \cite{Zhang,Chen,Larson}.  
However, the spontaneous 
emission rate of excited states used for optical transitions in experiments \cite{Colombe,Brennecke}
is much higher than the tunneling rate of the atoms between the two wells \cite{Albiez,Folling,Trotzky}.
Here we consider the two hyperfine states $|e\rangle$ and $|g\rangle$ of
${}^{87}$Rb with the transition frequency $2\pi\times{6.8}$~GHz \cite{Matthews}.
The coherence times \cite{Harber,Treutlein} 
of these hyperfine spin states ($|e\rangle$ and $|g\rangle$) are much longer 
than both the timescales of tunneling and atom-photon interactions.  
Therefore, this system offers possibilities 
for the study of how the tunnel couplings between the two spatially 
separated condensates affect the atom-photon dynamics.

We focus our investigation on the the system in the limits of the strong 
and weak tunnel couplings, respectively.  We find that the system has the 
different dark-state subspaces \cite{Fleischhauer} in these two tunneling regimes.  
In the weak-tunneling regime, the system has a family of dark states which can 
be used for producing quantum entanglement between the condensates.  
Here we propose to efficiently 
generate steady-state entanglement between the two spatially separated 
condensates by evolving to a mixture of dark states through the 
dissipation of the photon field \cite{Yang,Plenio,Joshi}.  Note 
that our scheme does not require any adjustment of the tunneling strength.  
It is different to other methods \cite{Ng1} which depend on the strength of tunnel
couplings to generate entanglement.  In addition, the entanglement generated between
the two condensates can be used for the implementation of quantum state transfer \cite{Bose}.
This may be useful for quantum information processing with atom-chip devices \cite{Treutlein}.

\begin{figure}[ht]
\centering
\includegraphics[height=3.0cm]{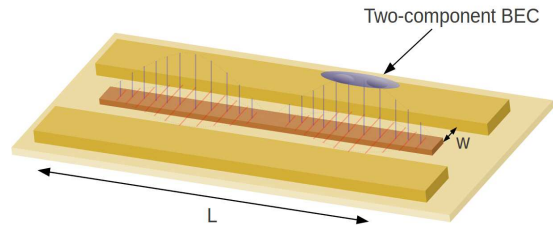}
\caption{ \label{fig1} (Color online) Schematic of a two-component BEC coupled to a single-mode of the photon field inside
a superconducting resonator. A two-component condensate is trapped in a double-well potential, and it is placed close to
the surface of the superconducting resonator.  The atoms are coupled to the magnetic field via their magnetic dipoles.
The parameters $L$ and $w$ are the length and width of the superconducting resonator, respectively. }
\end{figure}

This paper is organized as follows:  In Sec.~II, we introduce the system of a 
two-component condensate in a double-well potential, and the two-level atoms are
coupled to a superconducting resonator.  In Sec.~III, we derive the two 
effective Hamiltonians in the strong- and weak-tunneling regimes, respectively. 
In Sec.~IV, we investigate the dark-state subspaces and 
the atom-photon dynamics in the two tunneling limits.   In Sec.~V, we provide 
a method to produce the steady-state entanglement between the two condensates
in a double well. A summary is given in Sec.~VI.  In Appendix A, we discuss 
the validity of the effective Hamiltonian in the strong tunneling regime.

\section{System}
We consider a two-component BEC being trapped in a 
double-well potential \cite{Ng1}, and the condensate is placed near 
the surface of a superconducting resonator as shown in Fig.~\ref{fig1}.
The atoms, with two internal states $|e\rangle$ and $|g\rangle$, are coupled to 
a single mode of the photon field via their magnetic dipoles.  

\subsection{A two-component condensate trapped in a double-well potential}
We first introduce the system of a two-component condensate in
a one-dimensional (1D) double-well potential which can be described by
the Hamiltonian as
\begin{eqnarray}
%H_0&=&\sum_{\alpha}\int\!{dx}\Psi^\dag_{\alpha}(x)\Big[-\frac{\hbar^2}{2m_{\alpha}}\frac{\partial^2}{\partial{x^2}}+V_{{\rm DW}}(x)
%+{\tilde{U}_{\alpha}}\Psi^\dag_{\alpha}(x)\nonumber\\
%&&\times\Psi_{\alpha}(x)\Big]\Psi_{\alpha}(x)+2\tilde{U}_{eg}\int\!{dx}\Psi^\dag_e(x)\Psi^\dag_g(x)\nonumber\\
%&&\times\Psi_g(x)\Psi_e(x),\nonumber\\
H_0\!&=&\!\sum_{\alpha}\!\!\int\!{dx}\Psi^\dag_{\alpha}(x)\!\Big[\!-\frac{\hbar^2}{2m_{\alpha}}\frac{\partial^2}{\partial{x^2}}+V_{{\rm DW}}(x)
+{\tilde{U}_{\alpha}}\Psi^\dag_{\alpha}(x)\nonumber\\
&&\times\Psi_{\alpha}(x)\Big]\Psi_{\alpha}(x)\!+\!2\tilde{U}_{eg}\!\!\int\!\!{dx}\Psi^\dag_e(x)\Psi^\dag_g(x)\Psi_g(x)\Psi_e(x),\nonumber\\
\end{eqnarray}
where $\Psi_{\alpha}(x)$ is the field operator of the atoms for the internal state $|\alpha\rangle$ at the position $x$,
and the indices $\alpha=g,e$ represent the ground and the excited states, respectively.  Here $m_{\alpha}$ is 
the mass of the atom in the state $|\alpha\rangle$ and $V_{{\rm DW}}(x)$ is the 1D double-well potential
which is given by \cite{Maussang}
\begin{equation}
 V_{\rm DW}(x)=V_{d}\Big[1-\Big(\frac{x}{x_0}\Big)^2\Big]^2,
\end{equation}
where $V_d$ is the barrier height and $x_0$ is the distance between the two separate potential wells.  
The atoms are transversely confined in the $y$- and $z$-directions with the trap frequencies $\omega_{\perp}$.
The size of the ground-state wave function in the transverse motion is 
$a_{\perp}=\sqrt{\hbar/{m_{\alpha}\omega_{\perp}}}$ \cite{Olshanii,Pflanzer},
where $m_{e}$ and $m_{g}$ are nearly equal.  Since the transverse frequencies are much larger than the trap frequency in 
the $x$-direction, the transverse motions of the atoms are frozen out.
The parameters $\tilde{U}_{\alpha}$ and $\tilde{U}_{eg}$ are the effective 1D interaction strengths between the inter-, 
and the intra-component condensates, as \cite{Olshanii,Pflanzer}
\begin{eqnarray}
 \tilde{U}_{\alpha}&=&\frac{2\hbar^2{a_\alpha}}{m_{\alpha}a^2_{\perp}}\Big(1-C\frac{a_\alpha}{\sqrt{2}a_{\perp}}\Big)^{-1},\\
 \tilde{U}_{eg}&=&\frac{4\hbar^2m_em_g{a_{eg}}}{(m_{e}+m_{g})a^2_{\perp}}\Big(1-C\frac{a_{eg}}{\sqrt{2}a_{\perp}}\Big)^{-1},
\end{eqnarray}
where $C\approx{1.4603}$. The parameters $a_{\alpha}$ and $a_{eg}$ are the three-dimensional s-wave scattering lengths 
for the inter-, and the intra-component condensates.

We adopt the two-mode approximation \cite{Milburn} such that
the field operator $\Psi_\alpha(x)$ can be expanded in terms of the two localized mode functions 
$u_{\alpha_L}(x)$ and $u_{\alpha_R}(x)$ as,
\begin{eqnarray}
 \Psi_{\alpha}(x)&=&\alpha_{L}u_{\alpha{L}}(x)+\alpha_{R}u_{\alpha{R}}(x),
\end{eqnarray}
where $\alpha_{L}$ and $\alpha_{R}$ are the annihilator operators of the atoms in the state $\alpha=e,g$ 
for the left and right modes of the double-well potential, respectively.  The Hamiltonian of the system \cite{Ng1}, 
within the two-mode approximation, can be written as 
\begin{eqnarray}
\label{H0}
 H_0'&=&{\hbar}E_{e}(e^{\dag}_{L}e_{L}+e^\dag_{R}e_{R})+\hbar{E_g}(g^{\dag}_{R}g_{R}+g^{\dag}_{L}g_{L})\nonumber\\
&&-{\hbar}J_e(e^{\dag}_{L}e_{R}+e^\dag_{R}e_{L})-{\hbar}J_g(g^{\dag}_{L}g_{R}+g^\dag_{R}g_{L})\nonumber\\
&&+{\hbar}U_{ee}[(e^\dag_{L}e_{L})^2+(e^\dag_{R}e_{R})^2]+{\hbar}U_{gg}[(g^\dag_{L}g_{L})^2\nonumber\\
&&+(g^\dag_{R}g_{R})^2]+2{\hbar}U_{eg}(e^\dag_{L}e_{L}g^\dag_{L}g_{L}+e^\dag_{R}e_{R}g^\dag_{R}g_{R}),~~~~
\end{eqnarray}
where 
\begin{eqnarray}
E_{\alpha}&=&\frac{1}{\hbar}\int\!\!{dx}u^*_{\alpha{j}}(x)\Big[-\frac{\hbar^2}{2m_{\alpha}}\frac{\partial^2}{\partial{x^2}}+V_{\rm DW}(x)\Big]u_{\alpha{j}}(x),~~\\
J_{\alpha}&=&-\frac{1}{\hbar}\int\!\!{dr}u^*_{\alpha{L}}(x)\Big[-\frac{\hbar^2}{2m_{\alpha}}\frac{\partial}{\partial{x^2}}+V_{\rm DW}(x)\Big]u_{\alpha{R}}(x),~~~\\
U_{\alpha}&=&\frac{\tilde{U}_{\alpha}}{\hbar}\int\!\!{dx}|u_{\alpha{j}}(x)|^4,\\
U_{\alpha\beta}&=&\frac{\tilde{U}_{\alpha\beta}}{\hbar}\int\!\!{dr}|u_{\alpha{j}}(x)|^2|u_{\beta{j}}(x)|^2,
\end{eqnarray}
and $j=L,R$.
The positive parameters $E_\alpha$ and $J_\alpha$ \cite{tunparameter} are the ground-state frequencies of the localized mode $\alpha_{L,R}$, and the tunneling strengths between the two wells for the atoms in the states $\alpha$.  Here $U_\alpha$ and $U_{\alpha\beta}$ are the two positive parameters which describe the inter- and intra-component
interaction strengths, respectively.

\subsection{Atoms coupled to the photon field in a microwave cavity}
We consider that the atoms are coupled to a single-mode of the photon field via their magnetic dipoles \cite{Henschel}.  
Within the two-mode approximation, the Hamiltonian, describing the system of cavity field, the atoms and 
their interactions, is given by 
\begin{eqnarray}
 H_I&=&\hbar{\omega_a}a^\dag{a}+\hbar{\omega_0}(e^\dag_Le_L+e^\dag_Re_R)
+{\hbar}g[a(e^\dag_{L}g_{L}+e^\dag_{R}g_{R})\nonumber\\
&&+{\rm H.c.}],
\end{eqnarray}
where $\omega_a$ and $a$ are the frequency and the annihilator operator of the single-mode of the photon field,
and $\omega_0$ is the transition frequency of the two internal states.   
Here we have assumed that the wavelength of the microwave field (${\sim}~1$~cm) is much larger than the size of 
the condensate ($\sim~{10}~\mu$m) \cite{Schumm,Maussang}. Therefore, all atoms are coupled to the photon 
field with the same coupling strength 
$g=\mu_B\sqrt{\mu_0\omega_a/2\hbar{V}}$ \cite{Imamoglu}, where $\mu_B$ is the Bohr magneton, $\mu_0$ is the
vacuum permeability and $V$ is the volume of the superconducting resonator.  The coupling strength $g$
can attain $1$~kHz \cite{Imamoglu} if the volume $V$ of the superconducting resonator is taken as
$L\times{w}\times{{t_h}}\sim{1}~{\rm cm}\times{10}~\mu{\rm m}\times{200}~{\rm nm}$ \cite{Imamoglu,Wang},
where $L$ is the length, $w$ is the width, and ${t_h}$ is the thickness of the superconducting resonator.

\section{Effective Hamiltonians in strong and weak tunneling regimes: 
Low atomic excitations}
We will derive the effective Hamiltonians of the system in the limits of strong and weak tunnel couplings, 
respectively, where a few atomic excitations are only involved.  
Let us first write the total Hamiltonian of the system as
\begin{eqnarray}
\label{tHam}
 H&=&\hbar\omega_a{a^\dag{a}}+\hbar{\omega_0}(e^\dag_Le_L+e^\dag_Re_R)-{\hbar}J_e(e^{\dag}_{L}e_{R}+e^\dag_{R}e_{L})\nonumber\\
&&-{\hbar}J_g(g^{\dag}_{L}g_{R}+g^\dag_{R}g_{L})+{\hbar}U_{ee}[(e^\dag_{L}e_{L})^2+(e^\dag_{R}e_{R})^2]\nonumber\\
&&+{\hbar}U_{gg}[(g^\dag_{L}g_{L})^2+(g^\dag_{R}g_{R})^2]+2{\hbar}U_{eg}(e^\dag_{L}e_{L}g^\dag_{L}g_{L}\nonumber\\
&&+e^\dag_{R}e_{R}g^\dag_{R}g_{R})+{\hbar}g[a(e^\dag_{L}g_{L}+e^\dag_{R}g_{R})+{\rm H.c.}].
\end{eqnarray}
The total number of atoms $N$ is conserved. We have omitted the constant 
term $E_0N$ for a symmetric double well, where $E_{\alpha}\approx{E_0}$ 
for the two masses $m_e$ and $m_g$ being equal.   It is convenient to work in the rotating frame by applying the 
unitary transformation to the Hamiltonian $H$ in Eq.~(\ref{tHam}), where the unitary 
operator $U(t)$ is
\begin{eqnarray}
 U(t)&=&\exp{[-i\omega_a({a^\dag{a}}+e^\dag_Le_L+e^\dag_Re_R)t]}.
\end{eqnarray}
The transformed Hamiltonian becomes
\begin{eqnarray}
 H'&=&\hbar{\Delta}(e^\dag_Le_L+e^\dag_Re_R)-{\hbar}J_e(e^{\dag}_{L}e_{R}+e^\dag_{R}e_{L})-{\hbar}J_g(g^{\dag}_{L}g_{R}\nonumber\\
&&+g^\dag_{R}g_{L})+{\hbar}U_{ee}[(e^\dag_{L}e_{L})^2+(e^\dag_{R}e_{R})^2]+{\hbar}U_{gg}[(g^\dag_{L}g_{L})^2\nonumber\\
&&+(g^\dag_{R}g_{R})^2]+2{\hbar}U_{eg}(e^\dag_{L}e_{L}g^\dag_{L}g_{L}+e^\dag_{R}e_{R}g^\dag_{R}g_{R})\nonumber\\
&&+{\hbar}g[a(e^\dag_{L}g_{L}+e^\dag_{R}g_{R})+{\rm H.c.}]
\end{eqnarray}
where $\Delta=\omega_0-\omega_a$ is the detuning between the frequencies of 
the photon field and the two internal states.

In the strong tunneling regime, the tunnel coupling is dominant and the strength of 
atom-atom interactions is relatively weak.  On the contrary, in the weak tunneling regime, 
the atom-atom interactions become dominant and the tunneling strength is negligible.   
We will show that these two cases exhibit the different behaviours in
the atom-photon dynamics.  We will provide derivations of the two
effective Hamiltonians in the two tunneling limits in the 
following subsections.

\subsection{Strong-tunneling regime}
In the limit of the strong tunnel coupling,
the tunneling strengths are much larger than the strengths of the atom-atom interactions, i.e., 
$J_{e},J_{g}~{\gg}~U_{e},U_{g},U_{eg}$.  The total Hamiltonian of the system can be approximated as
\begin{eqnarray}
 H_1&=&\hbar\Delta(e^\dag_Le_L+e^\dag_Re_R)-{\hbar}J_e(e^{\dag}_{L}e_{R}+e^\dag_{R}e_{L})\nonumber\\
&&-{\hbar}J_g(g^{\dag}_{L}g_{R}+g^\dag_{R}g_{L})+{\hbar}g[a(e^\dag_{L}g_{L}+e^\dag_{R}g_{R})+{\rm H.c.}].\nonumber\\
\end{eqnarray}
We have neglected the terms of the atom-atom interactions in this Hamiltonian.

The symmetric and asymmetric modes $g_{\pm}$
and $e_{\pm}$ can be related to the localized
modes as
\begin{eqnarray}
 g_{\pm}&=&\frac{1}{\sqrt{2}}(g_L\pm{g_R}),\\
e_{\pm}&=&\frac{1}{\sqrt{2}}(e_L\pm{e_R}).
\end{eqnarray}
The Hamiltonian is then transformed as 
\begin{eqnarray}
\label{wtHam1}
 H_1'&=&{\hbar}(\Delta-J_e)e^\dag_+e_++\hbar({\Delta+J_e})e^\dag_-e_-
-{\hbar}J_g(g^{\dag}_{+}g_{+}\nonumber\\
&&-g^\dag_{-}g_{-})+{\hbar}g(ae^\dag_{+}g_{+}+{\rm H.c.})+{\hbar}g(ae^\dag_{-}g_{-}+{\rm H.c.}).\nonumber\\
\end{eqnarray}
Here the atoms are in symmetric (asymmetric) mode if they are populated in the states $g^k_+|0\rangle_+$ or $e^k_+|0\rangle_+$ 
($g^k_-|0\rangle_-$ or $e^k_-|0\rangle_-$), where $|0\rangle_+$ ($|0\rangle_-$) is
the vacuum state of the symmetric (asymmetric) mode and $k$ is a non-negative integer.

We consider the system to be initially prepared in the ground state 
in the limit of strong tunnel coupling,
i.e., the ground state of the symmetric mode.
The ground state can be obtained by applying the operator $(g^\dag_+)^N$ 
to the vacuum state $|0\rangle_+$ of the symmetric mode, i.e.,
\begin{equation}
\label{Psi_s}
 |\Psi_{1}(0)\rangle=\frac{1}{\sqrt{N!}}(g^{\dag}_+)^N|0\rangle_+,
\end{equation}
where $N$ is the total number of atoms.
Note that the atoms in the symmetric and asymmetric modes are independently coupled to 
the photon field in Eq.~(\ref{wtHam1}).  Therefore, all atoms in the symmetric mode are only involved in the 
dynamics of the atom-photon interactions if the system starts with the state $|\Psi_1(0)\rangle$ in
Eq.~(\ref{Psi_s}).  In fact, there are only a few excitations in the asymmetric
mode due to the atomic interactions.  
The effect of the excitations from the asymmetric 
mode to the dynamics of atom-photon interactions is very small.  
It is because the Rabi coupling strength cannot
be greatly enhanced with a small number of atoms in 
the asymmetric mode.  We briefly discuss the
validity of this assumption in Appendix A.

It is instructive to express the Hamiltonian in terms of angular momentum operators:
\begin{eqnarray}
\label{S+1}
S^{(+)}_+&=&g_{+}e^\dag_{+},\\
\label{S+2}
S^{(+)}_-&=&e_{+}g^\dag_{+},\\
\label{S+3}
S^{(+)}_z&=&\frac{1}{2}(e^\dag_{+}e_{+}-g^\dag_{+}g_{+}).
\end{eqnarray}
The Hamiltonian can be rewritten as 
\begin{eqnarray}
 \tilde{H}_1'&=&\hbar{\Delta}S^{(+)}_z+{\hbar}g(aS^{(+)}_++{\rm H.c.}).
\end{eqnarray}
For simplicity, we have assumed that the tunneling strengths $J_e$ and $J_g$
are equal. We also have omitted the constant term $\hbar{N}\Delta/2$.

By applying the Holstein-Primakoff transformation (HPT) \cite{Holstein}, the angular momentum operators 
can be mapped onto the harmonic oscillators which are given by
\begin{eqnarray}
\label{HPT1a}
 S^{(+)}_{+}&=&b^\dag\sqrt{N-b^\dag{b}},\\
\label{HPT1b} 
S^{(+)}_{-}&=&b\sqrt{N-b^\dag{b}},\\
\label{HPT1c} 
S^{(+)}_{z}&=&b^\dag{b}-\frac{N}{2}.
\end{eqnarray}
In the low degree of excitation, the mean excitation number $\langle{b^\dag{b}}\rangle$
are much smaller than the total number of atoms $N$.
The angular momentum operators can be approximated by the bosonic operators \cite{Ng1,Ng3}.  
The effective Hamiltonian can be obtained as
\begin{eqnarray}
\label{eHam1}
 H^{(1)}_{\rm eff}&=&\hbar{\Delta}b^\dag{b}+{\hbar}g\sqrt{N}(a{b^\dag}+{\rm H.c.}).
\end{eqnarray}
Note that the effective Rabi frequency is enhanced by a factor of $\sqrt{N}$.
This effective Hamiltonian $H^{(1)}_{\rm eff}$ in Eq.~(\ref{eHam1}) describes the interactions between the collective-excitation mode and the single mode of the photon field.

\subsection{Weak-tunneling regime}
Now we investigate the system in the weak tunneling regime, where the atom-atom interaction strengths
are much larger than the tunneling strengths, i.e, $U_e,U_g,U_{eg}{\gg}J_e,J_g$.
In this limit, we assume that the tunneling between the two condensates
is effectively turned off.  The total Hamiltonian can be approximated as
\begin{eqnarray}
 H_2&=&{\hbar}\Delta(e^\dag_Le_L+e^\dag_Re_R)+{\hbar}g[a(e^\dag_{L}g_{L}+e^\dag_{R}g_{R})+{\rm H.c.}]\nonumber\\
&&+{\hbar}U_{ee}[(e^\dag_{L}e_{L})^2+(e^\dag_{R}e_{R})^2]+{\hbar}U_{gg}[(g^\dag_{L}g_{L})^2\nonumber\\
&&+(g^\dag_{R}g_{R})^2]+2{\hbar}U_{eg}(e^\dag_{L}e_{L}g^\dag_{L}g_{L}+e^\dag_{R}e_{R}g^\dag_{R}g_{R}).
\end{eqnarray}
Here we have ignored the terms of the tunnel couplings.

This Hamiltonian can be expressed in terms of the angular momentum operators:
\begin{eqnarray}
S_{j+}&=&g_je^\dag_{j},\\
S_{j-}&=&e_{j}g^\dag_j,\\
S_{jz}&=&\frac{1}{2}(e^\dag_{j}e_{j}-g^\dag_{j}g_{j}),
\end{eqnarray}
where $j=L,R$.
Now the Hamiltonian is rewritten as
\begin{equation}
\label{wtHam2}
 \tilde{H}_2=\hbar\sum_{j=L,R}(\Delta+\delta)S_{jz}+{\hbar}g(aS_{j+}+{\rm H.c.})+\hbar\chi{S}^2_{jz},
\end{equation}
where $\delta=(U_{ee}-U_{gg})N/2$ and $\chi=U_{ee}+U_{gg}-2U_{eg}$.
We have omitted the constant term $\hbar(U_{ee}+U_{gg}+2U_{eg})N^2/16+\hbar{N}\Delta/2$ in Eq.~(\ref{wtHam2}).

We consider all atoms at the state $|g\rangle$ are 
initially prepared in the ground state of the Hamiltonian 
in Eq.~(\ref{wtHam2}),
which can be described by a product of two number states as
\begin{eqnarray}
\label{Psi_w}
 |\Psi_{2}(0)\rangle&=&|{N}/{2}\rangle_{g_L}|{N}/{2}\rangle_{g_R}.
\end{eqnarray}
Without loss of generality, we assume that $N$ is an even number.

We apply the HPT such that the angular momentum operators can be mapped onto
the harmonic oscillators as:
\begin{eqnarray}
S_{L+}&=&c^\dag\sqrt{N/2-c^\dag{c}},~~~~S_{L-}=c\sqrt{N/2-c^\dag{c}},\\
&&~~~~~~~~~~S_{Lz}=c^\dag{c}-\frac{N}{4},\\
S_{R+}&=&d^\dag\sqrt{N/2-d^\dag{d}},~~~~S_{R-}=d\sqrt{N/2-d^\dag{d}},\\
&&~~~~~~~~~~S_{Rz}=d^\dag{d}-\frac{N}{4},
\end{eqnarray}
If the mean numbers of the atomic excitations, $\langle{c^\dag{c}}\rangle$
and $\langle{d^\dag{d}}\rangle$, are much smaller than the number of atoms $N/2$
in each well, then the Hamiltonian can be approximated \cite{Ng1,Ng3} as
\begin{eqnarray}
\label{eHam2}
 H^{(2)}_{\rm eff}&=&\hbar\Delta_w(c^\dag{c}+d^\dag{d})+{\hbar}g\sqrt{\frac{N}{2}}[a(c^\dag+d^\dag)+{\rm H.c.}]\nonumber\\
&&+\hbar\chi[(c^\dag{c})^2+(d^\dag{d})^2],
\end{eqnarray}
where $\Delta_w=2\Delta+\delta-{\chi}N/2$.
The effective Rabi frequency is enhanced by a factor of $\sqrt{N/2}$.
The parameter $\chi$ is much smaller than the effective Rabi frequency because 
the scattering lengths of the inter- and intra-component condensates of ${}^{87}$Rb 
are very similar \cite{Matthews}.  We will ignore the terms with the parameter $\chi$ 
in Eq.~(\ref{eHam2}) in our later discussion.

The effective Hamiltonian $H^{(2)}_{\rm eff}$ in Eq.~(\ref{eHam2})
describes the interactions between the single mode of the photon
field and the two modes of the
collective excitations of the atoms in the left and right potential
wells, respectively.
This system can be described by a system of three coupled harmonic oscillators. 
The effective Rabi frequency for each atomic mode is proportional to the 
factor $\sqrt{N/2}$.  This is different to the effective Rabi frequency, in the 
strong-tunneling regime, which is proportional to the factor $\sqrt{N}$.

\section{Dark states and Quantum dynamics of the system}
\begin{figure}[ht]
\centering
\includegraphics[height=8.5cm]{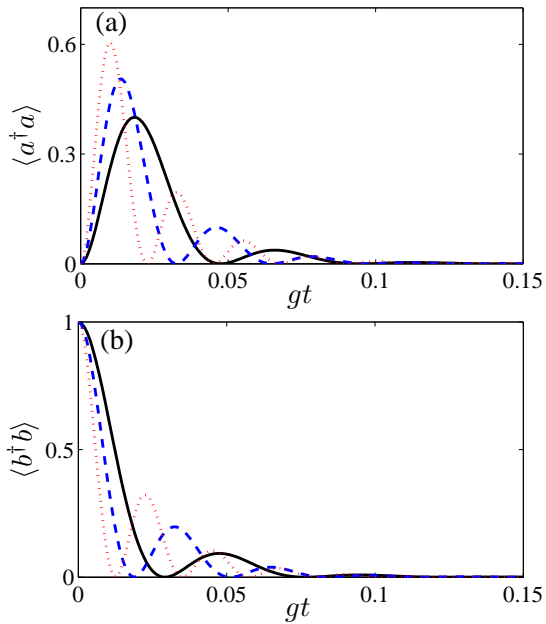}
\caption{ \label{fig_stnum} (Color online)
Time evolution of the mean photon number (a) and the mean atomic excitations (b)
with the damping rate $\kappa=100g$ and the detuning $\Delta=0$.  
The different number of atoms $N$ are $5\times{10^3}$ (black-solid line), $1\times{10^4}$ 
(blue-dashed line) and $2\times{10^4}$ (red-dotted line), respectively. 
}
\end{figure}
We now study dark states of the system which  
has different dark-state subspaces in the 
strong- and weak-tunneling regimes. 
Let us first introduce the definition of dark states.
Dark states \cite{Fleischhauer} are the eigenstates of the atom-photon 
interaction operator $\mathcal{V}$, with zero eigenvalues, i.e.,
\begin{eqnarray}
 \mathcal{V}|{\rm Dark}\rangle&=&0|{\rm Dark}\rangle,\\
&=&0.
\end{eqnarray}
Dark states, in the strong- and weak-tunneling regimes, in this system 
can be obtained as
\begin{eqnarray}
 H^{(j)}_{\rm eff}|D\rangle_j=0,
\end{eqnarray}
where $H^{(j)}_{\rm eff}$ are the two effective Hamiltonians in Eqs.~(\ref{eHam1}) and
(\ref{eHam2}) with zero detunings ($\Delta=\Delta_w=0$) and $j=1,2$.

In the limit of strong tunnel coupling, the dark state $|D\rangle_1$
is the product state of the vacuum state of the photon 
field and the ground state of the atomic mode $b$, which is given by
\begin{equation}
 |D\rangle_1=|0\rangle_a|0\rangle_b.
\end{equation}
This state is the ground state of the coupled system of the atoms
and the photon field.

\begin{figure}[ht]
\centering
\includegraphics[height=9.5cm]{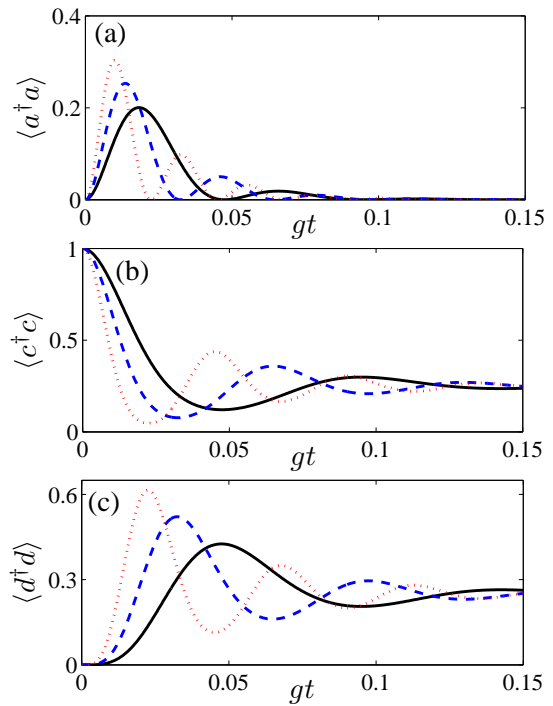}
\caption{ \label{fig_wtnum}  (Color online) 
Dynamics of the mean photon number and mean atomic excitation numbers
with the damping rate $\kappa=100g$ and the detuning $\Delta_w=0$.
(a) the mean photon number $\langle{a^\dag{a}}\rangle$ as a function
of the time $gt$.  Time evolution of the mean atomic excitation numbers of
the atomic mode $c$, in (b), and the atomic mode $d$, in (c) as shown.  
The different number of atoms $N$ are 
$5\times{10^3}$ (black-solid line), $1\times{10^4}$ (blue-dashed line) 
and $2\times{10^4}$ (red-dotted line), respectively.}
\end{figure}

In the weak-tunneling regime, the system has a family of dark states.  
The family of dark states are
\begin{equation}
\label{adark1}
 |D_{n}\rangle_{2}=|0\rangle_a|D^{a}_n\rangle,
\end{equation}
where 
\begin{equation}
\label{adark2}
 |D^{a}_n\rangle=2^{-n/2}\sum^n_{j=0}(-1)^j\sqrt{C^{n}_j}|n-j\rangle_c|j\rangle_d,
\end{equation}
and $C^{n}_j$ is the binomial coefficient.
The dark states $|D_{n}\rangle_2$ are the product state of the vacuum state $|0\rangle_a$ of
the photon field and the states $|D^{a}_n\rangle$ are the eigenstates of the operator $c+d$ with zero
eigenvalues. Note that 
the states $|D^{a}_n\rangle$ in Eq.~(\ref{adark2}) is a superposition of the states $|n-j\rangle_c|j\rangle_d$
which have the same degree of atomic excitations.

To gain more insight into dark states, 
let us first investigate the atom-photon dynamics subject to the 
dissipation of the photon field.  For a superconducting resonator with the frequency $\sim{40}$~GHz
can be cooled down to low temperatures ($\sim{25}$~mK) \cite{Hofheinz}.  
This allows us to consider the cavity field 
being weakly coupled to the reservoir at the zero temperature \cite{Ng4}. 
Note that the relaxation time (several $\mu$s) of the single photon inside the superconducting 
resonator is much shorter than the coherence time ($\sim{1}$s) of the cold atoms \cite{Harber,Treutlein}.  
The effect of the dissipation of the atoms caused by the noise of the surface of the 
superconductor is negligible \cite{Kasch}.  The main source of the dissipation is the damping of 
the photon field.  The dynamics of the system can be described by the master equation, 
for the zero temperature, as \cite{Ng4,Barnett}
\begin{eqnarray}
\label{dmaster} 
\dot{\rho}_j&=&-\frac{i}{\hbar}[H^{(j)}_{\rm eff},\rho_j]+\frac{\kappa}{2}(2a{\rho_j}a^\dag-a^\dag{a}\rho_j-\rho_j{a^\dag{a}}),
\end{eqnarray}
where $\rho_j$ is the density matrix of the total system, and $j=1,2$.
Obviously, the dark states $|D\rangle_1$ and $|D_n\rangle_2$ are the steady-state solutions of
the master equation in Eq.~(\ref{dmaster}).  Thus, the dark states are robust against the
dissipation of the photon field.
In the strong tunneling regime, the steady state 
is the dark state $|D\rangle_1$.  In the weak tunneling regime, 
the state of the condensates evolves as a mixture of dark states $|D_n\rangle_2$ 
through the dissipation of the photon field.

Now we study the dynamics of the system in the strong-tunneling
regime, where the state is prepared as $|0\rangle_a|1\rangle_b$ and $|1\rangle_b$
is a number state.  We plot the time of evolution of the mean photon number
and mean atomic-excitation number in Fig.~\ref{fig_stnum}.  The 
mean photon number and mean atomic excitations undergo a few oscillations and then 
both of them decay to zero.  We also see that the faster rate of 
oscillations can be obtained if a larger number of atoms $N$ are used.

We proceed to investigate the atom-photon dynamics in the weak-tunneling regime.
The system is initially prepared as the state $|0\rangle_a|1\rangle_c|0\rangle_d$,
where $|1\rangle_c$ is a number state.  In Fig.~\ref{fig_wtnum}, we plot
the mean photon number, and the mean excitation numbers of the two atomic modes
versus the time.  When the atom-photon interactions are turned on, the 
excitation number of the atomic mode $c$ decreases while the mean photon number
increases as shown in Fig.~\ref{fig_wtnum}.  Afterwards, the mean excitation number of 
the atomic mode $d$ starts to increase.
This means that the energy of the atomic mode $c$ transfers to the photon
field and the atomic mode $d$ absorbs the energy from the photon field.
In this way, the two atomic modes exchange the energy via the photon field.
The faster rate of exchanging energy between the atoms and the photon
field can be attained if a larger number of atoms $N$ are used.
We also note that the mean photon number in Fig.~\ref{fig_wtnum}(a) 
is about half of the mean photon number in Fig.~\ref{fig_stnum}(a).  It is because
the atoms in the atomic mode $d$, in the weak tunneling regime, absorbs the 
energy from the photon field.

In Fig.~\ref{fig_wtnum}(a), the mean photon number decays to zero after 
a period of time.  However, the mean excitation numbers of modes $c$
and $d$ remain non-zero as shown in Figs.~\ref{fig_wtnum} (b) and (c).  
It is because the state of the atoms evolves to
a mixture of dark  states $|D_0\rangle_2$ and $|D_1\rangle_2$, and 
a single excitation is shared by the atoms in the dark state $|D_1\rangle_2$.
This results in the non-zero excitation numbers of the two atomic modes.

\section{Generation of entanglement between two spatially separated condensates}
\begin{figure}[ht]
\centering
\includegraphics[height=8.5cm]{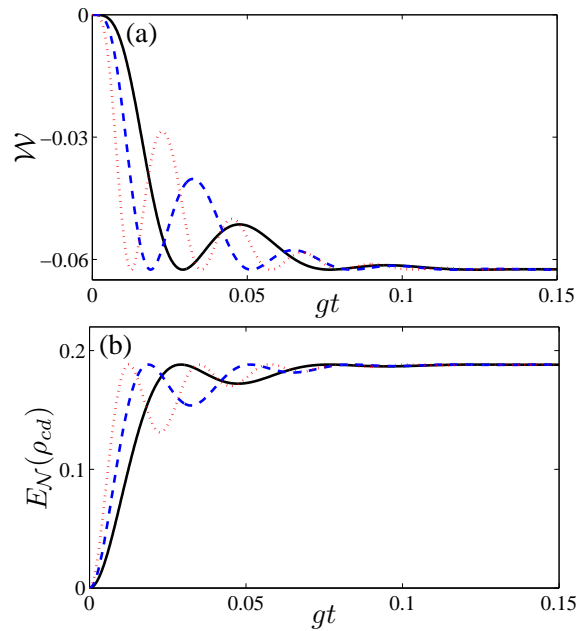}
\caption{ \label{fig_wtent1} (Color online) Time evolution of the entanglement witness 
in (a) and logarithmic negativity in (b), for the damping rate $\kappa=100g$ and the detuning
$\Delta_w=0$.  The different number of atoms $N$ are $5\times{10^3}$ (black-solid line), 
$1\times{10^4}$ (blue-dashed line) and $2\times{10^4}$ (red-dotted line), respectively.}
\end{figure}
We have shown that the system has the different dark-state
subspaces in the two tunneling limits.
Now we study the entanglement between the condensates in the
two different potential wells in the weak tunneling regime.  
In this regime, the system has a family of dark states which can 
be used for generating entanglement.
Here we consider the tunneling between the wells to be effectively turned off.  
Therefore, the two independent condensates in the two potential wells are 
initially unentangled.  We will show that steady-state entanglement between 
the two condensates can be produced by evolving to a mixture of dark 
states $\{|D_n\rangle_2\}$ through the dissipation of the photon field 
\cite{Yang,Plenio,Joshi}.

To study the quantum entanglement between the two atomic modes
$c$ and $d$, it is necessary to obtain the density matrix of the 
atomic condensate.  By tracing out the system of the photon field,
we can obtain the density matrix $\rho_{cd}$,
\begin{equation}
 \rho_{cd}={\rm Tr}_{a}(\rho),
\end{equation}
where $\rho$ is the density matrix of the total system.
Let us first examine the entanglement of a single dark state $|D_n\rangle_2$.
For a dark state $|D_n\rangle_2$ in Eq.~(\ref{adark1}), 
the density matrix $\rho_{cd}$ is given by
\begin{equation}
\rho_{cd}=|D^a_n\rangle\langle{D}^a_n|,
\end{equation}
where $|D^a_n\rangle$
is the state in Eq.~(\ref{adark2}).
The degree of entanglement between the two atomic modes can be 
quantified by the von Neumann entropy.
It is defined as
\begin{eqnarray}
E_F&=&-{\rm Tr}(\rho_{c}\ln\rho_{c}), 
\end{eqnarray}
where $\rho_{c}={\rm Tr}_d(\rho_{cd})$ 
is the reduced density matrix. 
\begin{figure}[ht]
\centering
\includegraphics[height=8.5cm]{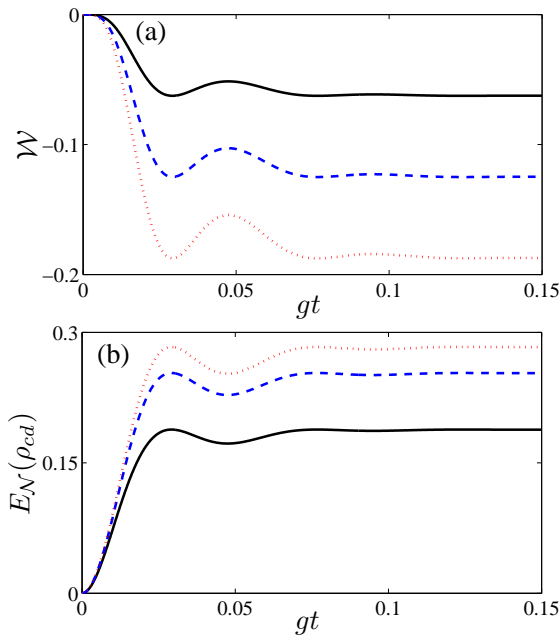}
\caption{ \label{fig_wtent2} 
(Color online) Plot of the dynamics of entanglement.
(a) entanglement witness $\mathcal{W}$ and 
(b) logarithmic negativity $E_{\mathcal{N}}(\rho_{cd})$
as a function of the time $gt$. The initial state 
$|0\rangle_a|n\rangle_c|0\rangle_d$ with 
the different excitation numbers $n$ are shown, for $n=1$ (black-solid line), $n=2$ 
(blue-dashed line) and $n=3$ (red-dotted line), respectively.
The parameters are $\kappa=100g$, $\Delta_w=0$ and $N=5{\times}{10^3}$. }
\end{figure} 
The von Neumann entropy is 
\begin{equation}
\label{vNentropy}
 E_F=-2^{-n}\sum^n_{j=0}C^{n}_j\ln|{2^{-n}C^n_j}|.
\end{equation}
Thus, the state $|D^a_n\rangle$ is an entangled state. 
The degree of two-mode entanglement becomes higher for 
larger $n$.

In general, this density matrix $\rho_{cd}$ is a mixed state.
To quantify the entanglement of a mixed state, the logarithmic 
negativity can be used \cite{Vidal}.
The definition of the logarithmic negativity is \cite{Vidal}
\begin{eqnarray}
 E_{\mathcal{N}}(\rho_{cd})&=&\log_2\parallel{\rho^{T_c}_{cd}}\parallel,
\end{eqnarray}
where $\rho^{T_c}_{cd}$ is the partial transpose of the density matrix $\rho_{cd}$
and $\parallel{\cdot}\parallel$ is the trace norm.

However, the logarithmic negativity is difficult to be experimentally
determined.  It is very useful to study an 
experimentally accessible quantity to detect the quantum entanglement
between the two bosonic modes \cite{Hillery}.  
If an inequality  
\begin{eqnarray}
 |\langle{cd^\dag}\rangle|^2>\langle{n_c}{n_d}\rangle,
\end{eqnarray}
is satisfied \cite{Hillery}, then the state is an entangled state.
Here $n_c=c^\dag{c}$ and $n_d=d^\dag{d}$ are the number operators of the atomic modes
$c$ and $d$, respectively.
For convenience, this quantity $\mathcal{W}$ is defined as
\begin{equation}
\mathcal{W}=\langle{n_c}{n_d}\rangle-|\langle{cd^\dag}\rangle|^2.
\end{equation}
If $\mathcal{W}$ is negative, then the state is non-separable. 
This quantity $\mathcal{W}$ is called as an entanglement witness \cite{Horodecki}.

We investigate the dynamics of entanglement between the two atomic modes.  We
consider an initial state as a product state of the three modes, i.e., 
$|0\rangle_a|1\rangle_c|0\rangle_d$, where $|1\rangle_c$ is a number state.  We plot
the entanglement witness and logarithmic negativity versus time as shown
in Fig.~\ref{fig_wtent1}.  This figure shows that the entanglement witness
decreases and logarithmic negativity increases with a similar rate, 
and then they saturate after a short time.  This
shows that the steady-state entanglement can be produced in a short time via
the dissipative photon field.
The entanglement can also be produced faster if a larger number of atoms are used.
Besides, we can see that the entanglement witness is consistent with the
logarithmic negativity to indicate the degree of entanglement.  
The entanglement witness is a faithful indicator for detecting the entanglement
between the two bosonic modes.

Next, we study the generation of entanglement by using an initial state 
$|0\rangle_a|n\rangle_c|0\rangle_d$ with a higher degree of excitation, 
where $|n\rangle_c$ is a number state and 
$n$ is larger than one.  In Fig.~\ref{fig_wtent2}, the entanglement witness and 
logarithmic negativity are plotted versus the time.
It shows that a higher degree of the entanglement can be obtained if higher
excitation numbers $n=2,3$ are used.

\vspace*{-0.1cm}
\section{Summary}
\vspace*{-0.1cm}
We have studied a two-component condensate 
in a double-well potential, where the atoms are 
magnetically coupled to a single-mode of the photon 
field inside a superconducting resonator.  The system has the different 
dark-state subspaces in the strong- and weak-tunneling regimes, respectively,
and it gives rises to the different dynamics of atomic excitations in the two regimes.
Steady-state entanglement between the two spatially separated condensates 
can be produced by evolving to a mixture of dark states through the dissipative 
photon field.  We have shown that the entanglement can be faithfully
indicated by an entanglement witness.

\vspace*{0.2cm}
\begin{acknowledgments}
\vspace*{0.2cm}
H.T.N. thank David Hallwood for his careful reading 
and helpful comment, and C. K. Law for his useful discussion. 
This work was partially supported by 
U.S. National Science Foundation. We also
would like to acknowledge the partial support of National
Science Council of Taiwan (Grant No. 97-2112-M-002-003-MY3) 
and National Taiwan University (Grant No. 99R80869).
\end{acknowledgments}

\vspace*{0.1cm}
%\newpage
\appendix
\section{Validity of the effective Hamiltonian in the strong-tunneling regime}
\begin{figure}[ht]
\centering
\includegraphics[height=4.0cm]{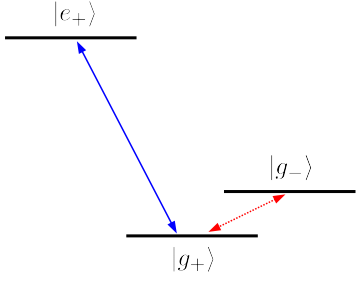}
\caption{ \label{app_2_level1} (Color online) Level scheme of the atoms in the double-well potential.  In 
the strong-tunneling regime, the atoms, with the two states $|e_+\rangle$ and $|g_+\rangle$, 
in the symmetric mode are coupled to the cavity field.  The two ground states of
the symmetric and asymmetric modes ($|g_+\rangle$ and $|g_-\rangle$) 
are coupled to each other via the atom-atom interactions.}
\end{figure}
In this appendix, we examine the validity of the effective Hamiltonian $H^{(1)}_{\rm eff}$ 
in Eq.~(\ref{eHam1}) in the limit of strong tunnel coupling.
We express the Hamiltonian in term of the symmetric-mode and asymmetric-mode operators as:
\begin{eqnarray}
\label{wtHamf}
\tilde{H}
&=&{\hbar}(\Delta-J_e)e^\dag_+e_++\hbar(\Delta+J_e)e^\dag_-e_--{\hbar}J_g(g^{\dag}_{+}g_{+}\nonumber\\
&&-g^\dag_{-}g_{-})+{\hbar}g(ae^\dag_{+}g_{+}+{\rm H.c.})
+{\hbar}g(ae^\dag_{-}g_{-}+{\rm H.c.})\nonumber\\
&&+\frac{{\hbar}U_{ee}}{2}\big[(e^\dag_+e_++e^\dag_-e_-)^2+(e^\dag_+e_-+e^\dag_-e_+)^2\big]\nonumber\\
&&+\frac{{\hbar}U_{gg}}{2}\big[(g^\dag_+g_++g^\dag_-g_-)^2+(g^\dag_+g_-+g^\dag_-g_+)^2\big]\nonumber\\
&&+{{\hbar}U_{eg}}\big[(e^\dag_+e_++e^\dag_-e_-)(g^\dag_+g_++g^\dag_-g_-)+(e^\dag_+e_-\nonumber\\
&&+e^\dag_-e_+)(g^\dag_+g_-+g^\dag_-g_+)\big].
\end{eqnarray}

Let us define 
\begin{eqnarray}
 F^+&=&g^\dag_-g_+,~~~~F^-=g^\dag_+g_-,\\
F_3&=&\frac{1}{2}(g^\dag_-g_--g^\dag_+g_+).
\end{eqnarray}
The commutation relations $[F_3,F_{\pm}]=\pm{F}_{\pm}$ and $[F_+,F_-]=2F_{z}$ are satisfied.
The operators $F_{\pm}$ and $F_z$, and $S^{(+)}_{\pm}$ and $S^{(+)}_z$ in Eqs.~(\ref{S+1}) to (\ref{S+3}) generate a SU(3) algebra \cite{Jin}.
In the limit of large $N$, we can apply the HPT to the operators:
\begin{eqnarray}
F^+&=&f^\dag\sqrt{N-f^\dag{f}},~~~~F^-=f\sqrt{N-f^\dag{f}},\\
&&~~~~~~F_3=f^\dag{f}-N/2.
\end{eqnarray}
Assume that the mean excitation number $\langle{f^\dag{f}}\rangle$ is much 
smaller than $N$, we can approximate the operators as \cite{Jin}:
\begin{eqnarray}
 f^\dag&=&\frac{1}{\sqrt{N}}F^+,~~~~f=\frac{1}{\sqrt{N}}F^-.
\end{eqnarray}
In the low-degree-of-excitation regime, the approximated Hamiltonian can be written as
\begin{eqnarray}
 \tilde{H}&\approx&{\hbar}\omega_a{a^\dag{a}}+\hbar{\omega'}b^\dag{b}+{\hbar}g\sqrt{N}(a{b^\dag}+{\rm H.c.})+\tilde{H}'.\nonumber\\
\end{eqnarray}
\begin{figure}[ht]
\centering
\includegraphics[height=6cm]{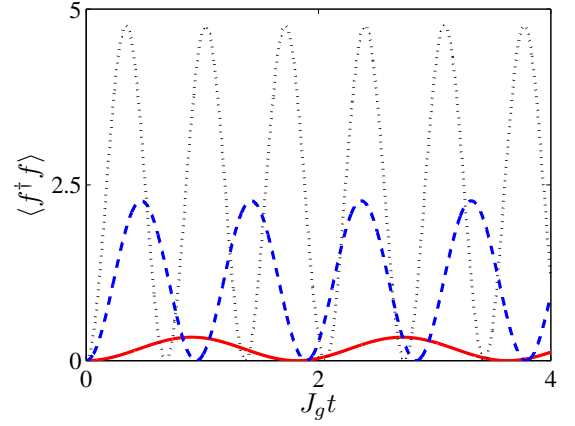}
\caption{ \label{appfig1} (Color online) The expectation value $\langle{f^\dag{f}}\rangle$ 
as a function of the time $J_g{t}$. Different strengths of $U_{gg}N$ are shown: 
$U_{gg}N=J_g$ (red-solid line), $U_{gg}N=5{J_g}$ (blue-dashed line), and 
$U_{gg}N=10{J_g}$ (black-dotted line).}
\end{figure}
The Hamiltonian $\tilde{H}'$, contains the terms of the operators in the asymmetric mode and the terms from the nonlinear interactions,
and the constant terms are omitted, which can be written as 
\begin{eqnarray}
\tilde{H}'&=&{\hbar}J_gf^\dag{f}+\frac{{\hbar}U_{gg}N}{2}(f^\dag+f)^2
+{\hbar}(\Delta+J_e)(e^\dag_-e_-\nonumber\\
&&-e^\dag_-e_-)+{\hbar}g(ae^\dag_{-}g_{-}+{\rm H.c.})+\frac{{\hbar}U_{ee}}{2}\big[(e^\dag_+e_+\nonumber\\
&&+e^\dag_-e_-)^2+(e^\dag_+e_-+e^\dag_-e_+)^2\big]+\frac{{\hbar}U_{gg}}{2}(g^\dag_+g_+\nonumber\\
&&+g^\dag_-g_-)^2+{\hbar}U_{eg}\big[(e^\dag_+e_++e^\dag_-e_-)(g^\dag_+g_++g^\dag_-g_-)\nonumber\\
&&+\sqrt{N}(e^\dag_+e_-+e^\dag_-e_+)(f^\dag+f)\big].
\end{eqnarray}
Here we consider the number of atoms in the excited states to be very small.  We also
assume that the strength of the Rabi coupling $g$ is weak compared to the tunneling strength
$J_g$ and nonlinear strength $U_{gg}N$ but $g$ is much stronger than $U_{ee},U_{gg}$ and $U_{eg}$.  
Therefore, the Hamiltonian $\tilde{H}'$ can be approximated by the Hamiltonian $H''$ as
\begin{eqnarray}
\label{squeezeH}
 \tilde{H}''&=&\hbar\lambda_1f^\dag{f}+{\hbar}\lambda_2(f^{\dag2}+f^2),
\end{eqnarray}
where
\begin{eqnarray}
 \lambda_1&=&J_g+U_{gg}N,\\
 \lambda_2&=&\frac{U_{gg}N}{2}.
\end{eqnarray}
From Eq. (\ref{squeezeH}), nonlinear interactions can give rise to the transitions of the atoms in the symmetric mode
to the atoms in the asymmetric mode, and vice versa.
The level scheme is shown in Fig.~\ref{app_2_level1}.
Note that this Hamiltonian $\tilde{H}''$ is exactly solvable.  The time-evolution operator can be factorized as \cite{Barnett}
\begin{eqnarray}
S(t)\!&=&\!\exp(-i\tilde{H}''t/\hbar),\\
&=&\!\exp\!\Big(\frac{\Lambda_2}{2}{f}^{\dag{2}}\!\Big)\!\exp\!\Big[\frac{\ln(\Lambda_1)}{4}(f^\dag{f}+{f}f^\dag)\!\Big]\!\exp\!\Big(\frac{\Lambda_2}{2}f^2\!\Big),\nonumber\\
\end{eqnarray}
where 
\begin{eqnarray}
 \Lambda_1&=&\Big(\cosh\beta-\frac{\lambda'_1}{2\beta}\sinh\beta\Big)^{-2},\\
\Lambda_2&=&\frac{2\lambda_2'\sinh\beta}{2\beta\cosh\beta-\lambda_1'\sinh\beta},\\
\beta^2&=&\frac{{\lambda'_1}^2}{4}-{\lambda'_2}^2,\\
\lambda_1'&=&-2i\lambda_1t,~~\lambda_2'=-2i\lambda_2t.
\end{eqnarray}

We then apply the time-evolution operator $S(t)$ to the vacuum state $|0\rangle_{f}$ of
the mode $f$.  The state becomes 
\begin{equation}
 |\Psi_s(t)\rangle=\Lambda^{1/4}_1\sum^{\infty}_{n=0}\frac{\sqrt{(2n)!}}{n!}\Big(\frac{\Lambda_2}{2}\Big)^{n}|2n\rangle_{f}.
\end{equation}
The mean excitation number $\langle{f^\dag{f}}\rangle$ is
\begin{equation}
 \langle{f^\dag{f}}\rangle=\big|\Lambda^{1/4}_1\big|^2\sum^{\infty}_{n=0}\frac{{n(2n)!}\Lambda^{2n}_2}{2^{2n-1}(n!)^2}.
\end{equation}
In Fig.~\ref{appfig1}, we plot the expectation value $\langle{f^\dag{f}}\rangle$ versus the time,
for the different strengths of atomic interaction $U_{gg}N$.  It is shown that there are
only a few excitations in the asymmetric mode even if the atomic-interaction strength $U_{gg}N$
is much larger than the tunneling strength $J_g$. 
Therefore, the Rabi coupling
strength cannot be greatly increased due to the collective
enhancement.

\end{document}